\def\preprint{0}                
\def\preprint{1}                
\def\comment#1{}
\if\preprint1
        \documentclass[usenatbib]{mn2e}
        \usepackage{times}
        \usepackage{graphicx}
        \usepackage{calc}

\else
        \documentstyle[astrop-bib,referee,times]{mn}
        \newcommand{\includegraphics}[1]{}
\fi

\newcommand{\mum}{\ifmmode{\rm \mu m}\else{$\mu$m}\fi}
\def\oversim#1#2{\lower0.5pt\vbox{\baselineskip0pt \lineskip-0.5pt
     \ialign{$\mathsurround0pt #1\hfil##\hfil$\crcr#2\crcr\sim\crcr}}}
\def\aap{{\rm A\&A}}
\def\mnras{{\rm MNRAS}}
\def\apj{{\rm ApJ}}
\def\apjl{{\rm ApJl}}
\def\aj{{\rm AJ}}
\def\apjs{{\rm ApJS}}
\def\pasp{{\rm PASP}}
\def\apss{{\rm ApSS}}

\title[The dust and gas content of carbon stars toward the Galactic Halo]{Dust and gas in 
carbon stars toward the Galactic Halo}
\author[E. Lagadec et al.]{Eric Lagadec$^{1}$\thanks{E-mail:elagadec@eso.org},
  G.~C.~Sloan$^2$,  Albert~A.~Zijlstra$^3$, 
  Nicolas Mauron$^4$, J.R. Houck$^2$\newauthor \\
$^1$European Southern Observatory, Karl Schwarzschildstrasse 2, 
  Garching 85748, Germany\\
$^2$Department of Astronomy, Cornell University, 
  222 Space Sciences Building, Ithaca NY 14853-6801, USA\\ 
$^3$Jodrell Bank Center for Astrophysics, Alan Turing Building, 
  School of Physics and Astronomy The University of Manchester, 
  Oxford Street, Manchester, M13 9PL, UK \\
$^4$Laboratoire Univers et Particules de Montpellier
UMR 5299 CNRS/IN2P3 - UM2
Universite Montpellier 2
Place Eugene Bataillon
34095 Montellier , France\\
}
\pubyear{2008}

\begin{document}

\date{Accepted . Received}

\pagerange{\pageref{firstpage}--\pageref{lastpage}} \pubyear{2002}

\maketitle

\label{firstpage}


\begin{abstract}

We present {\it Spitzer} IRS spectra of four carbon stars located in the
Galactic Halo and the thick disc.  The spectra display typical features of
carbon stars with SiC dust emission and C$_2$H$_2$ molecular absorption.  Dust
radiative transfer models and infrared colors enable us to determine the dust
production rates for these stars whilst prior CO measurements yield
expansion velocities and total mass-loss rates.

The gas properties (low expansion velocities (around 7 km/s) and strong C$_2$H$_2$ molecular
absorption bands) are consistent with the stars being metal-poor.  However
the dust content of these stars (strong SiC emission bands) is very similar to what is observed in metal-rich carbon stars.
The strong SiC emission  may indicate that the carbon stars derive
from a metal-rich population, or that these AGB stars produce silicon.

The origin of the halo carbon stars is not known.  They may be extrinsinc  halo
 stars belonging to the halo population,  they may have been accreted from
a satellite galaxy such as the Sagittarius Dwarf Spheroidal Galaxy,
or they may be escapees from the galactic disk. If the stars are intrinsically
metal-rich, an origin in the disc would be most likely.  If an $\alpha$-element
enhancement can be confirmed, it would argue for an origin in the halo (which
is known to be $\alpha$-enhanced) or a Galactic satellite.

\end{abstract}



\section{Introduction} 

Low- and intermediate-mass stars (LIMS, 0.8--8~M$_{\odot}$ on 
the main sequence) end their lives with an intense mass-loss 
episode on the asymptotic giant branch (AGB).  The mass loss 
from AGB stars is one of the main sources of enrichment of 
the interstellar medium (ISM) with newly synthesized 
elements.  In our Galaxy, mass loss from AGB stars produces 
well over half of the measured dust \citep{geh89}.  The
AGB also dominates the measurable dust production in the LMC 
and SMC \citep{mat09,sri09,boy12}.

The mass-loss mechanism is not fully understood, but a 
two-step mechanism is widely accepted as plausible (e.g., Winters et al. 2000). Pulsation 
from the star extends the atmosphere, where the material
is dense and cool enough to form dust.  Radiation pressure 
drives the mass loss, accelerating the grains while the gas
is carried along by friction.  Dust formation is the key 
step in the mass-loss process.  Its composition depends on 
the C/O ratio. If C/O\,$>$1, i.e. the star is carbon-rich, then most of the oxygen is 
trapped in volatile CO molecules, which have high dissociation energy. 
The dust  forms from the remaining refractory elements, and consists  of 
amorphous carbon and SiC.  For C/O\,$<$1, the dust will be 
oxygen-rich, consisting of silicates and aluminium oxides..

Understanding the effect of metallicity on the mass loss is 
of prime importance to determining the yields of LIMS in 
different galaxies and understanding the formation of dust in 
the early Universe.
  We thus carried out several surveys of
mass-losing AGB stars in the Local Group with the {\it 
Spitzer} Infrared Spectrograph \citep[IRS][]{hou04}, 
targeting the Large Magellanic Cloud \citep[LMC][]{zij06,
mat06,slo08}, Small Magellanic Cloud \citep[SMC][]{slo06,
lag07,slo08}, and several dwarf spheroidal galaxies
\citep{mat07,lag09,slo09,slo12}.

The results show that dust-production rates from oxygen-rich AGB stars are 
lower in more metal-poor environments, but for carbon stars,
metallicity shows no strong influence \citep{gro07,slo08}.
Subtantial dust-production rates are observed for carbon stars with 
metallicities as low as $\sim$0.1 Z$_{\odot}$ \citep{lag08,
slo09,slo12}.  


Most of the material ejected via the mass-loss process is in 
the gaseous form.  The gas mass in the envelope of an AGB 
star is typically two orders of magnitude higher than the 
dust mass to which our infrared surveys are
sensitive.  To fully characterize the 
mass loss from AGB stars and its dependence on
metallicity, one needs to study the {\it gas} mass loss in 
environments with different metallicities.  

Carbon stars have
recently been discovered in the Galactic Halo. These  are thought 
to be metal-poor \citep{ti98, mau04,mau05,mau07}. 
\cite{gro97} observed the CO emission from a carbon star in
the Halo.  We recently observed a sample of carbon Halo stars 
in the CO J $= 3 \rightarrow 2$ transition and found that 
their expansion velocities were low compared to AGB stars in
the Galactic disc with similar infrared colors \citep{lag10}.

Here, we report new {\it Spitzer} IRS observations of four of 
these carbon stars.  We place these observations in the 
context of our CO observations and additional IRS 
observations of three other likely carbon Halo stars.

\section{The sample} 

\cite{lag10} observed six carbon stars in the Halo using the 
James Clerk Maxwell Telescope (JCMT) to study the CO J$=3 
\rightarrow 2$ transition at 345 GHz.  Those stars were
selected from catalogs of carbon stars in the Halo by 
\cite{ti98}, \cite{mau04,mau05,mau07}, and \cite{mau08}.
Each target had to be a bright source as observed by the 
{\it Infrared Astronomical Satellite} ({\it IRAS}) and have
a $J-K$ color of $\sim$3--4.  The near-infrared color
requirement ensured that all of the dust shells had similar 
optical depths.

We selected five of the six JCMT targets as part of an IRS 
program for the final months of the cryogenic portion of the 
{\it Spitzer} mission.  We obtained spectra for four of our 
sample before the cryogens were exhausted, with the fourth 
spectrum coming on 2009 May 15, the day that {\it Spitzer} 
ran out of liquid helium.  In fact, this spectrum (of 
IRAS~08427) was the penultimate spectrum ever obtained by 
the IRS.

In this paper we also examine three additional carbon stars 
likely to be in the Halo.  \cite{slo10} published the IRS 
spectrum of Lyng{\aa}~7~V1 as part of their sample of AGB 
stars in globular clusters.  Carbon stars are rare in
globular clusters, making it possible that this star is in 
the foreground or background of the cluster.  However, all 
estimates of its distance coincided with the position of the
cluster.  Lyng{\aa}~7~V1 could be either a carbon Halo star 
in the immediate vicinity of Lyng{\aa}~7, or it could 
actually be a member.  \cite{lag09} published the spectra of 
several carbon stars in the Sagittarius Dwarf Spheroidal galaxy (Sgr dSph)
(previously known as  the Sagittarius Dwarf Elliptical 
Galaxy (SagDEG)), including IRAS 18384$-$3310 and IRAS 19074$-$3233 (herafter
IRAS 18383 and IRAS 19074).  The 
radial velocities of these two sources, as determined from 
optical spectra, showed that they were not members of Sgr dSph, 
and distance estimates placed them at least 10 kpc in the
foreground.

Table\,\ref{targets} gives the observational properties of 
the nine probable carbon Halo stars observed with either JCMT
or the IRS.  For the six Halo stars with JCMT observations,
the classifications of their membership in the thick disk,
Halo, or Sagittarius Stream are from \cite{lag10}, who used 
their radial velocities, distances, and galactic coordinates as
indicators.  The last two targets in the table, IRAS~18384
and IRAS~19074, are likely to be in the Halo, based on their
radial velocities ($-$23 and $-$73 km/s, respectively).
Lyng{\aa}~7~V1 is only 400 pc from the Galactic plane, 
consistent with membership of the old disc or thick disc.  We will
revisit these assignments in Sec.\,5.1 below.  

\begin{table*}
\caption[]{\label{targets} The observed sample of carbon 
stars in the Galactic Halo.  The near-infrared photometry is 
from 2MASS, except for Lyng{\aa}~7~V1, which is from 
\cite{slo10}.}
\begin{center}
\begin{tabular}{lllrlllrrrll}
\hline
Name & RA & Dec & \multicolumn{2}{c}{Gal.\ coordinates} & $K_s$ & $J-K_s$ & F$_{12}$ 
          & Distance & Scale height & Possible   & {Observations} \\
          & \multicolumn{2}{c}{(J2000)} & l$^{II}$ & b$^{II}$ & (mag) & (mag) & (Jy) & (kpc) & (kpc) & population & \\
\hline

IRAS 04188$+$0122 & 04 21 27.25 & $+$01 29 13.4 & 192.2 & $-$32.0 & 6.42 & 3.28 &  3.37   &  6.5 & $-$3.4 & Thick Disc & JCMT/IRS \\
IRAS 08427$+$0338 & 08 45 22.27 & $+$03 27 11.2 & 223.5 & $+$26.8 & 6.26 & 3.41 &  6.50   &  5.5 &    2.5 & Thick Disc & JCMT/IRS \\
IRAS 11308$-$1020 & 11 33 24.57 & $-$10 36 58.6 & 273.7 & $+$47.8 & 4.57 & 3.83 & 57.37   &  2.5 &    1.9 & Thick Disc & JCMT\\
IRAS 12560$+$1656 & 12 58 33.50 & $+$16 40 12.0 & 312.3 & $+$79.4 & 7.82 & 3.48 &  0.77   & 12.0 &   11.8 & Sgr Stream & JCMT \\
Lyng{\aa} 7 V1    & 16 11 02.05 & $-$55 19 13.5 & 328.8 & $-$02.8 & 7.25 & 4.10 & 2.99    &  7.3 & $-$0.4 & Thick Disc & IRS \\
IRAS 16339$-$0317 & 16 36 31.5  & $-$03 23 35.0 &  12.7 & $+$27.8 & 6.10 & 3.91 & 14.57   &  4.9 &    2.3 & Halo       & JCMT/IRS \\
IRAS 18120$+$4530 & 18 13 29.6  & $+$45 31 17.0 &  73.1 & $+$25.3 & 6.71 & 3.81 &  7.86   &  6.7 &    2.9 & Halo       & JCMT/IRS \\
IRAS 18384$-$3310 & 18 41 43.50 & $-$33 07 16.6 &   1.9 & $-$12.6 & 8.58 & 3.00 &  0.80   & 15.0 & $-$3.3 & Halo       & IRS \\
IRAS 19074$-$3233 & 19 10 39.87 & $-$32 28 37.3 &   5.0 & $-$17.9 & 8.35 & 3.61 &  1.73   & 14.5 & $-$4.5 & Halo       & IRS \\
\hline \\
\end{tabular}
\end{center}
\end{table*}



\begin{figure*}
\begin{center}
\includegraphics[width=18cm]{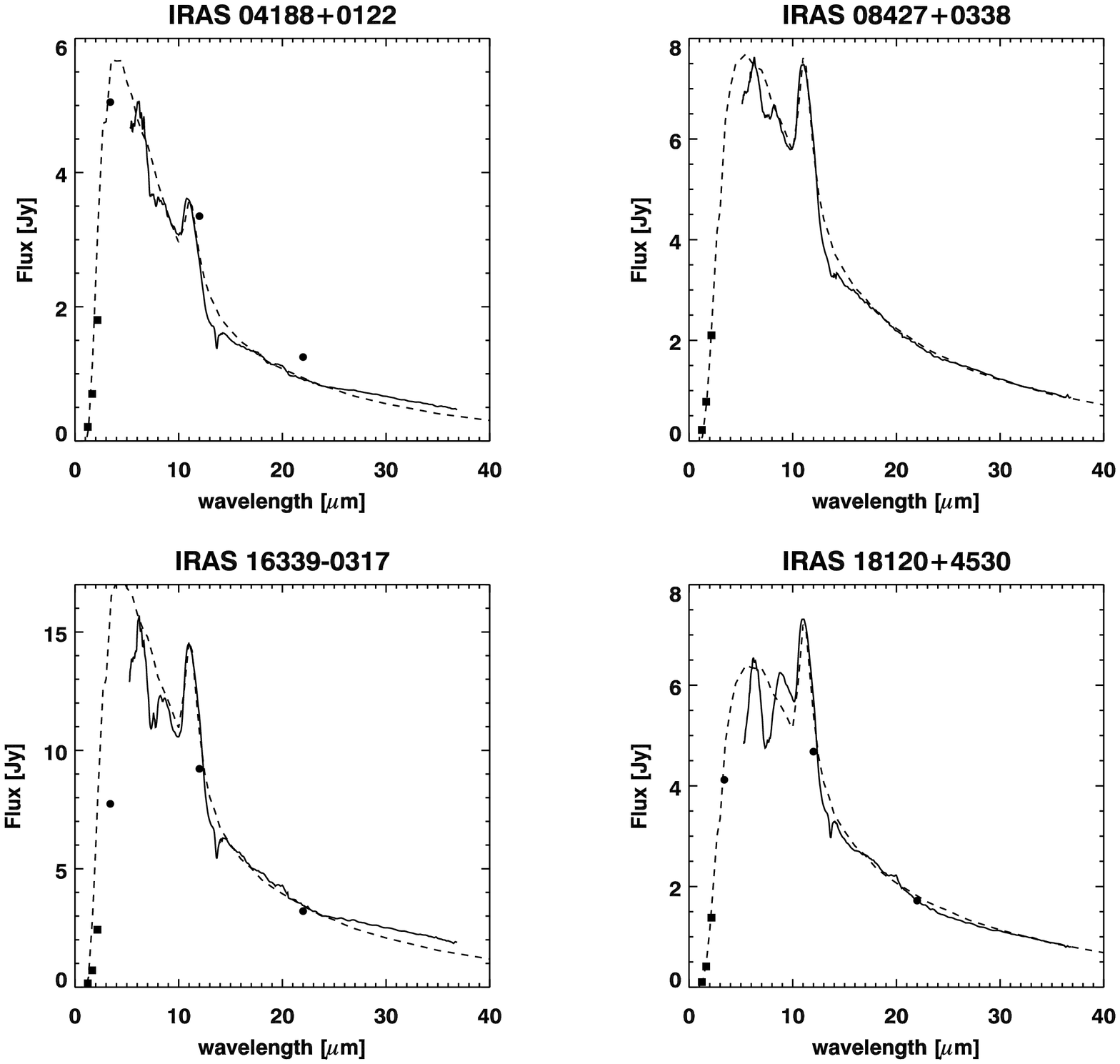}
\caption{\label{spitz}{\it Spitzer} IRS spectra of the four 
Halo carbon stars we observed.  Filled squares and circles 
are 2MASS JHK$_s$ and WISE photometry.  The dashed line 
represents the best fitting radiative transfer model.}
\end{center}
\end{figure*}
 

All IRS observations used both the Short-Low (SL) and 
Long-Low (LL) modules to cover the wavelength range 
5--37~\mum\ at resolutions of $\sim$100.  Before extracting
spectra from the IRS images, we (1) differenced them to 
remove background emission and additive spikes and divots, 
and (2) replaced pixels suspected of multiplicative spikes 
and divots with values calculated from neighboring pixels.
We used the standard tapered-column extraction algorithm
available with CUPID to obtain the spectra from the
images.\footnote{CUPID is the Customizable User Pipeline for
IRS Data, and it is available from the website of the {\it
Spitzer} Science Center.}  
When combining spectra from the two nod positions in each 
aperture, we flagged and ignored bad data which appeared in 
only one of the nods.  The spectra were then corrected with 
scalar multiplication to remove discontinuities between 
spectral segments.  These corrections were generally between 
5 and 10\%, but for IRAS 04188, the two SL segments were 
raised by 21 and 13\% (for the 5--7.5 and 7.5--14~\mum\ 
segments, respectively), probably due to a slight 
mispointing.  The spectral flux standards were HR~6348 (K0 
III) in SL and HR~6348 and HD~173511 (K5 III) in LL.  For 
more information, we refer the reader to the more detailed 
description of the observing method and data reduction by 
\cite{slo12}.\footnote{The only difference between their 
methodology and ours is that the brightness of our sample did 
not require the use of  the optimal extraction algorithm 
introduced by \cite{leb10}.}

 
\section{Description of the spectra}  
\label{spectra} 

%

Fig.\,\ref{spitz} displays the four infrared spectra obtained 
in our observing program, along with photometry from 2MASS
\citep{cut03} and the Wide-field Infrared Survey Experiment
(WISE \cite{wri10}).  The spectra confirm all four sources
as dusty carbon stars.  Each shows a prominent SiC dust
emission feature at 11.3~\mum, as well as absorption bands
from acetylene (C$_2$H$_2$) at 7.5 and 13.7~\mum.  The
absorption band only partially covered by the IRS at 
5~\mum\ is due to CO and is often strong in carbon stars.
The continuum between these various features and bands is
primarily due to emission from amorphous carbon grains.
 Some sources also display a broad 30 $\mu$m feature. \footnote { We will refer to this feature as the "MgS" feature,
although the reader should be aware of the continuing debate about
whether or not MgS produces the 30-$\mu$m feature.  See, for example,
Hony et al. (2002), Zhang et al. (2009) and Lombaert et al. (2012).}

To analyze these spectra, we apply the Manchester Method,
which was introduced by \cite{slo06} and \cite{zij06}.  This
method determines two colors  from the spectra in four 
narrow bands.  The [6.4]$-$[9.3] color is a good estimate of
the optical depth of the warm dust, while the [16.5]$-$[21.5] 
color provides an estimate of the overall dust temperature. 
For all the observed stars, we measured the strength of the 
spectral features from SiC dust at 11.3~\mum\ and acetylene
gas at 7.5 and 13.7~\mum\ using line segments to estimate
the continuum.  In each case, we defined the continuum using
wavelength intervals to either side of the feature.  We also
measured the strength of the MgS dust emission at 
$\sim$30~\mum, but because the red edge of the feature is
outside the IRS spectral range, we used a blackbody with  
the temperature derived from the [16.5]$-$[21.5] color to 
extrapolate the continuum under this dust feature.  
The strength of the dust features are reported as the ratio 
between the integrated flux in the feature and the continuum 
underneath it.  For the SiC feature, we also report its
integrated flux, scaled to a distance of 10 pc.  For the 
acetylene bands, we report the equivalent widths.  
Table\,\ref{ew.dat} presents all of these results.

To compare the strength of the observed features with ones in 
environments with different metallicities, we used a control 
sample of stars observed with the IRS and the SWS on the {\it
Infrared Space Observatory} ({\it ISO}.  This sample includes
spectra from AGB stars in the Galaxy, LMC, SMC, Sgr dSph, and
Fornax (see the references in Sec.\,1).  In all of the plots 
presented here, we use open symbols for Galactic stars. The 
symbols from largest to smallest represent stars in the Halo,
the Thick Disc, and the Galactic SWS sample, respectively. 
 Uncertainties were determined when coadding spectra extracted from 
individual images and propagated through all subsequent measurements.
The uncertainties in the strengths of the feature in the IRS spectra
include the effect of the uncertainty in the fitted continuum.

\begin{figure*}
\begin{center}
\includegraphics[width=18cm]{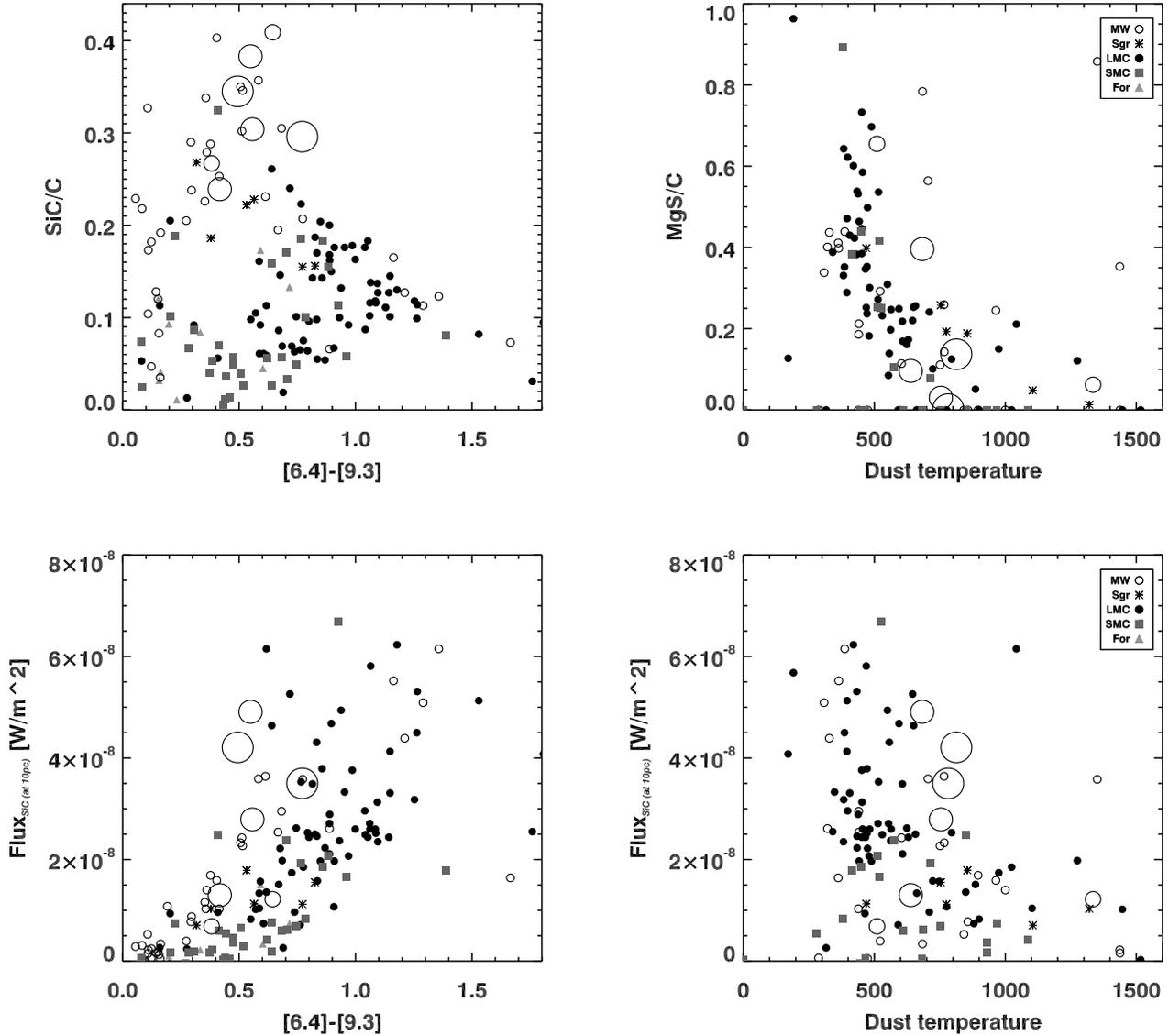}
\caption{\label{dust}Left:  Strength of the SiC dust emission 
feature as a function of the [6.4]$-$[9.3] color, with the 
top panel showing the strength normalized to the underlying 
continuum and the bottom panel showing the integrated flux 
normalized to a distance of 10 pc.  
Right:  Strength of the dust features as a function of dust
temperature (as estimated from the [16.5]$-$[21.5] color).
The large open circles are the seven stars from the present 
study (the largest ones represent Halo stars, while the 
intermediate ones represent thick disc stars).  The small
open circles and other symbols are from the comparison
samples, as shown in the inset keys.}
\end{center}
\end{figure*}

Fig.\,\ref{dust} shows the strength of the SiC and MgS dust
features as a function of the [6.4]$-$[9.3] color and the
temperature derived from the [16.5]$-$[21.5] color.  The 
strength of the MgS feature is low for all the observed 
stars, most likely due to the relatively high dust 
temperatures of the Halo stars.  MgS starts to form around 
600~K and completes its formation around 300\,K \citep{nuth85}. 
Fig.\,\ref{dust} shows that the dependence of the SiC feature on the dust temperature 
is less clear.  SiC tends to be weaker at high temperature 
(between 1000 and 1500~K) than at low temperature (between 500 
and 1000~K).
 
The upper left panel in Fig.\,\ref{dust} shows that the 
Galactic carbon stars follow one sequence in SiC/continuum 
strength as a function of [6.4]$-$[9.3] color, while the 
stars in the LMC, SMC, and Fornax follow another.
\cite{lag07} explained these two sequences as a consequence 
of differences in the condensation of amorphous carbon dust 
at different metallicities. The Halo carbon stars 
in our sample follow the same sequence as the Galactic carbon
stars, indicating a similar dust formation sequence and perhaps
a similar metallicity as well.  

The lower left panel in Fig.\,\ref{dust} shows  the 
integrated fluxes of the SiC features, normalized to a 
common distance of 10 pc.  While the separation between the
two sequences is much smaller compared to the upper left
panel, it is still apparent.  Again, the Halo stars in
our sample tend to follow the Galactic sequence.  This plot
removes the influence of the amorphous carbon continuum on
the measured strength of the SiC feature and shows that in
absolute terms, the feature is weaker in more metal-poor
carbon stars.

\begin{figure*}
\begin{center}
\includegraphics[width=18cm]{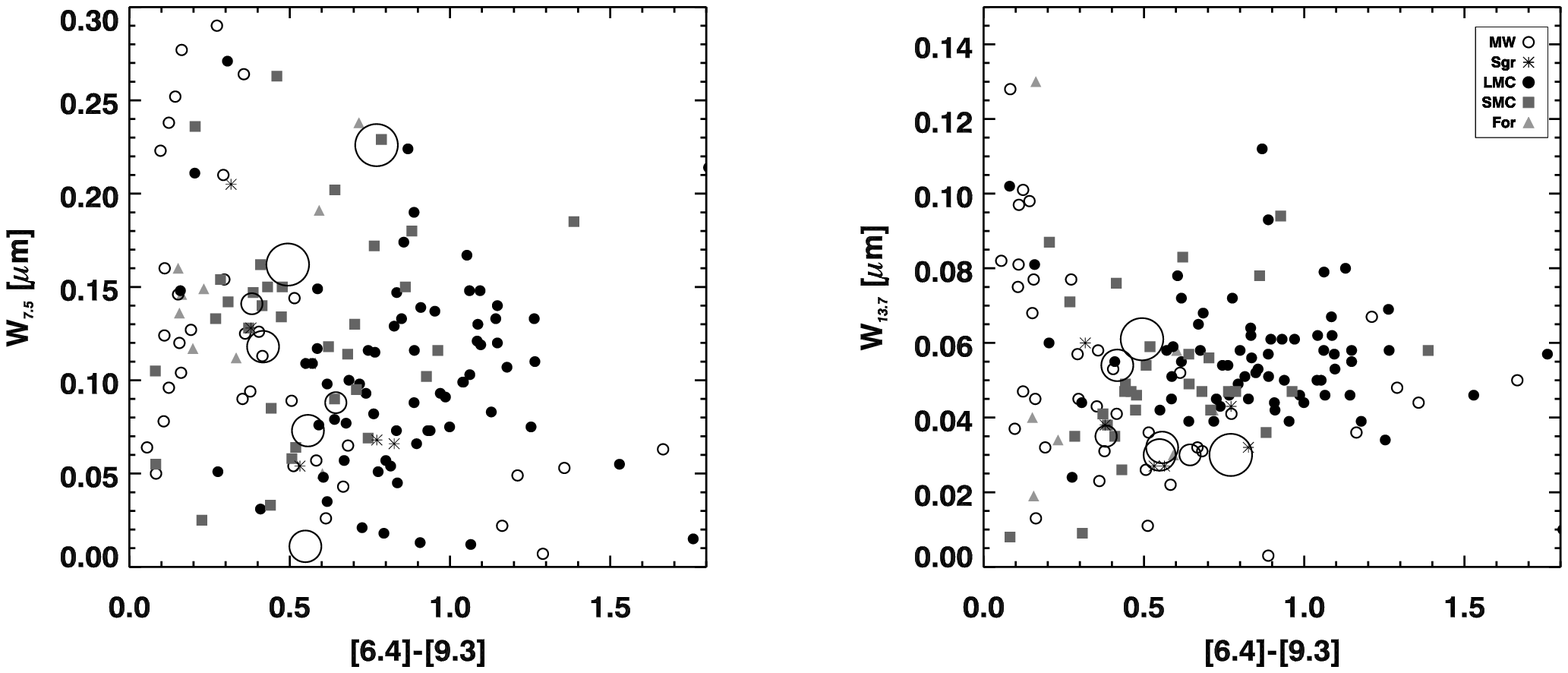}
\caption{\label{gas}Left:  Equivalent width  of the 7.5~\mum\ 
C$_2$H$_2$ feature as a function of the [6.4]$-$[9.3] color. 
Right:  Equivalent width of the 13.7~\mum\ C$_2$H$_2$ feature 
as a function of the [6.4]$-$[9.3] color.  The symbols are as
defined in Fig.\,\ref{gas}.}
\end{center}
\end{figure*}

Fig.\,\ref{gas} displays the equivalent width of the 7.5 and 
13.7~\mum\ C$_2$H$_2$ bands as a function of [6.4]$-$[9.3] 
color.  In our small sample, the Halo stars have stronger 
7.5~\mum\ acetylene bands than the stars assigned to the 
thick disc.  To compare these stars to the other populations,
it is necessary to limit the comparison to stars with similar
[6.4]$-$[9.3] colors, since these have similar column densities of
amorphous carbon dust around them.  The two Halo stars show
band strengths much like those in metal-poor environments 
like the SMC or Fornax, while the two thick disc stars are
more similar to carbon stars in the LMC and the Galaxy.
The comparisons of the 13.7~\mum\ feature are less 
instructive, because we only measure the narrow Q branch of 
the feature, which is saturated for high column densities.

Most of the spectroscopic studies based on IRS data 
referenced in this paper have noted the general tendency for
the 7.5~\mum\ acetylene band to strengthen in more metal-poor
samples.  \cite{mat07} suggest that this trend is the result
of higher C/O ratios in the outflows from the stars, but
\cite{slo12} have suggested it could result from  less
efficient dust formation.  If high-temperature condensates 
such as TiC act as seeds for production of amorphous carbon
grains from acetylene gas, the lower abundance of heavy 
elements like Ti in metal-poor carbon stars would lead to an
excess of carbon-rich gas to carbon-rich dust.

\begin{table*} 
\caption[]{\label{ew.dat}  Spectral measurements for our 
sample:  The two colors, the equivalent widths (EW)of the 
acetylene bands, the SiC dust emission strength (normalized
to the continuum and normalized to 10 pc), the MgS dust
emission strength (normalized to the continuum), and the dust
temperature, derived from the [16.5]$-$[21.5] color.}
\begin{center} 
\begin{tabular}{lccccccccllllllll} 
\hline 
                &                 &                 &  EW (7.5~\mum)  & EW (13.7~\mum)  &                 & SiC flux at 10 pc      &                 &  T \\
target          & [6.4]$-$[9.3]   & [16.5]$-$[21.5] &  (\mum)         & (\mum)          & SiC/cont.       & (10$^{-8}$ W m$^{-2}$) & MgS/cont.       & (K) \\ 
\hline 
IRAS 04188      & 0.417$\pm$0.009 & 0.212$\pm$0.011 & 0.118$\pm$0.011 & 0.054$\pm$0.008 & 0.239$\pm$0.007 & 1.30$\pm$0.04          & 0.096$\pm$0.014 &  639 $\pm$ 29 \\ 
IRAS 08427      & 0.557$\pm$0.007 & 0.176$\pm$0.012 & 0.073$\pm$0.005 & 0.032$\pm$0.003 & 0.304$\pm$0.005 & 2.79$\pm$0.05          & 0.030$\pm$0.015 &  754 $\pm$ 45 \\ 
Lynga{\aa} 7 V1 & 0.549$\pm$0.007 & 0.197$\pm$0.013 & 0.011$\pm$0.004 & 0.030$\pm$0.003 & 0.383$\pm$0.009 & 4.91$\pm$0.11          & 0.396$\pm$0.016 &  683 $\pm$ 39 \\ 
IRAS 16339      & 0.494$\pm$0.010 & 0.163$\pm$0.015 & 0.162$\pm$0.005 & 0.061$\pm$0.005 & 0.345$\pm$0.006 & 4.21$\pm$0.08          & 0.137$\pm$0.019 &  813 $\pm$ 65 \\ 
IRAS 18120      & 0.771$\pm$0.007 & 0.170$\pm$0.014 & 0.226$\pm$0.006 & 0.030$\pm$0.003 & 0.296$\pm$0.004 & 3.50$\pm$0.04          & 0.002$\pm$0.017 &  782 $\pm$ 54 \\ 
IRAS 18384      & 0.382$\pm$0.011 & 0.273$\pm$0.010 & 0.141$\pm$0.003 & 0.035$\pm$0.003 & 0.267$\pm$0.005 & 0.69$\pm$0.01          & 0.655$\pm$0.013 &  510 $\pm$ 16 \\ 
IRAS 19074      & 0.644$\pm$0.005 & 0.097$\pm$0.016 & 0.088$\pm$0.003 & 0.030$\pm$0.002 & 0.409$\pm$0.006 & 1.22$\pm$0.02          & 0.062$\pm$0.020 & 1335 $\pm$188 \\ 
              
\hline \\ 
\end{tabular} 
\end{center} 
\end{table*} 

\section{Dust-production rates and gas mass-loss rates} 

\subsection{Dust-production rates} 

To determine the dust-production rate (DPR, or $\dot{D}$) from
our carbon stars, we follow the approach of \cite{lag10}, who 
use two techniques.  The first method uses near- and
mid-infrared flux of the stars \citep{whi94,whi06,lag08}.
\cite{whi06} give the following relation:
\begin{eqnarray}
\label{mass-loss}
\nonumber \log(\dot{M}_{\rm total}) &&=-7.668+0.7305(K-[12])\\
\nonumber  &&-5.398 \times10^{-2}(K-[12])^2  \\
  &&+1.343\times10^{-3}(K-[12])^3,
\end{eqnarray}
where $K$ is the 2MASS $K_s$ magnitude, and [12] is the IRAS 
12~\mum\ magnitude, assuming a zero-magnitude flux of 28.3 Jy. 
We estimate the DPR from the total mass-loss rate by dividing
by the assumed gas-to-dust ratio of 200.  The relation 
obtained by \cite{whi06} assumes an expansion velocity of 
19~km/s.  We scale our results using the expansion velocity as
measured by the JCMT.

The second method uses radiative transfer models.  Here, we
present new models using the {\it Spitzer} spectra as an
additional constraint that was not available when \cite{lag10} 
generated the earlier models.  We used the radiative transfer 
code DUSTY \citep{ive99}, as described by \cite{lag10}.  This 
code solves the one-dimensional radiative transport problem in 
a dusty environment.  All of the models assume that the
irradiation comes from a point source (the central star) at 
the center of a spherical dusty envelope.  The circumstellar 
envelope is filled with material from a radiatively driven 
wind.  All of the stars are carbon-rich, and the dust consists 
of amorphous carbon and SiC.  Optical properties 
for these dust grains are taken from \cite{han88} and 
\cite{peg88}, respectively.  The grain size distribution is 
taken as a typical MRN distribution, with a grain size $a$ 
varying from 0.0005 to 0.25~\mum\ distributed according to a 
power law with n($a$)$\propto$$a^{-q}$, where $q$=3.5 
\citep{mat77}. We are aware that this grain size distribution 
was estimated for the interstellar medium, where dust grains
are thought to be significantly processed, but this is as good an
assumption as any given the lack of 
information about dust grain sizes in these stars. The outer radius of the dust shell was set to 
10$^3$ times the inner radius; this parameter has a negligible 
effect on our models.

To model the emission from the central star, we used a 
hydrostatic model including molecular opacities \citep{loi01,
gro07}.  Our aim was to fit the spectral energy distribution, 
defined by the 2MASS photometry ($J$, $H$ and $K_s$) and our
{\it Spitzer} data to estimate the DPRs.  Our models included 
only dust, so we do not attempt to fit the molecular 
absorption bands seen in the {\it Spitzer} spectra.  We fixed 
the dust temperature at the inner radius to be 1200~K and the 
effective temperature of the central star, T$_{\rm eff}$ = 
2800~K, unless a satisfactory fit could not be obtained with 
these parameters.

DUSTY gives total (gas+dust) mass-loss rates assuming a 
gas-to-dust ratio of 200.  The DPR can be obtained by 
dividing by this value, assuming that the gas and dust have
the same expansion velocities.  The expansion velocity is an 
output for the DUSTY models, and we scaled our results using 
the expansion velocities measured with JCMT  and assumed it to be 10 km.s$^{-1}$
when no measurements were available.

Table\,\ref{per_ml} gives the results from the fitted DUSTY
models.  Note that where available, the DPRs determined from
the $K_s$$-$[12] color are simply taken from \cite{lag10}.

  
\begin{table*} 
\begin{center} 
\caption[]{\label{per_ml} Dust production rates (DPRs) and 
results from radiative transfer models for the observed 
stars and some comparison stars.  For the Halo stars, these
data are from \cite{lag10}, except the last two entries,
which are in the Halo between SagDEG and the Galactic Center.
Those data are from \cite{lag09}.  The modelling for 
Lyng{\aa}~7~V1 is new, based on data presented by 
\cite{slo10}.}

\begin{center} 
\begin{tabular}{cccccccccccccccccccrlllllll} 
\hline 
Target    & Luminosity & $DPR_{\rm DUSTY}$ & T$_{eff}$ & T$_{\rm in}$ & SiC/AmC
          & $\tau$ (11$\mu$m) & $DPR_{\rm col}$& Distance \\ 
          & (10$^3$L$_\odot)$   & (10$^{-8}$M$_{\odot}$yr$^{-1}$) & (K)& (K)&
          &$\times$10$^{-1}$ & (10$^{-8}$M$_{\odot}$yr$^{-1}$) & kpc \\ 
\hline 
IRAS 04188     &  8.0 & 0.8 & 2650 & 1700 & 0.10  & 2.3 & 1.0     &  6.5 \\ 
IRAS 08427     &  7.5 & 2.3 & 2800 & 1200 & 0.13  & 2.5 & 2.2     &  5.5 \\ 
IRAS 11308     & 16.8 & 2.9 & 2800 & 1200 & 0.00  & 1.6 & 2.4     &  2.5 \\ 
IRAS 12560     &  7.5 & 0.3 & 2800 & 1200 & 0.00  & 1.1 & 0.2     & 12.0 \\ 
Lyng{\aa} 7 V1 &  9.0 & 3.3 & 2800 & 1200 & 0.15  & 2.6 & 2.1     &  7.3 \\
IRAS 16339     & 13.9 & 1.2 & 2800 & 1400 & 0.14  & 2.3 & 1.8     &  4.9 \\  
IRAS 18120     &  8.0 & 1.3 & 2800 & 1100 & 0.15  & 2.6 & 1.3     &  6.7 \\ 
IRAS 18384     & 11.5 & 2.6 & 2800 & 1200 & 0.10  & 1.1 & 0.9     & 14.5 \\
IRAS 19074     & 13.3 & 3.8 & 2800 & 1200 & 0.20  & 2.1 & 1.3     & 15.0 \\
\hline \\ 

\end{tabular} 
\end{center} 
\end{center} 
\end{table*} 

\subsection{Gas mass-loss rates} 

\begin{table*}
\caption[]{\label{line_values} CO J=3--2 lines properties of 
the observed Halo stars, gas mass-loss rates and gas-to-dust 
mass ratio determined using radiative transfer models and 
infrared colors.}
\begin{center}
\begin{tabular}{llllllllllllllll}
\hline
Target     & $v_{\rm exp}$ & $I_{CO}$ & $\dot{M}_{gas}$ & $\psi_D$& $\psi_c$ \\
           & km\,s$^{-1}$  & K\,km\,s$^{-1}$ & 10$^{-6}$M${_{\odot}}$yr$^{-1}$ & & \\
\hline
IRAS 04188 & 11.5 &  0.77 &  6.8 & 846 &  677 \\
IRAS 08427 & 16.5 &  0.55 &  4.2 & 184 &  193 \\
IRAS 11308 & 11.5 & 10.86 & 13   & 456 &  551 \\
IRAS 12560 &  3.0 &  0.14 &  2.6 & 866 &  1300 \\
IRAS 16339 &  8.5 &  2.59 &  11 & 905 & 603 \\ 
IRAS 18120 &  6.5 &  0.72 & 5.4   & 415 &  415 \\
\hline \\

\end{tabular}
\end{center}
\end{table*}
 
Spectroscopic observations of the CO lines are our best means of
determining the gas mass-loss rates from AGB stars.  CO is, 
after H$_2$, the most abundant circumstellar molecule \citep{olo97} 
and can be used easily as a probe of the circumstellar 
medium.  Since the 1980s, numerous works, combining such 
observations and radiative transfer models, have attempted to 
find a formula linking the CO line properties and the gas 
mass-loss rates \citep[e.g.][]{km85}.  From this point on, we
will equate the total mass-loss rate with that determined from
the gas.

\cite{ram08} presented a reliable and up-to-date formula to 
determine the total mass-loss rate from CO observations:
\begin{equation}
\dot{M}=S_J(I_{CO}\theta{_b^2}D^2)^{a_j}v{_e^{b_j}}f{_{CO}^{-C_j}},
\end{equation} 
where $I_{CO}$, $D$, $V_e$ and $f_{CO}$ are the CO line 
intensity (in K\,km\,s$^{-1}$), the distance (in kpc), the 
expansion velocity (in km\,s$^{-1}$) and the CO abundance, 
respectively.  The parameters $S_J$, $a_J$, $b_J$ and $c_J$ 
depend on the CO transition observed.  For the CO 
J\,$=$\,3$\rightarrow$\,2 line, they are 3.8$\times$10$^{-11}$, 
0.91, 0.39 and 0.45, respectively.  $\theta{_{mb}}$ is the full width 
at half maximum (FWHM) of 
the main beam of the telescope, which is 14 arcsec at JCMT.  
 We adopted the standard solar abundance  of 8.78$\times$10$^{-4}$ for f$_{CO}$ (Asplund et al. (2009)).  This value
could be lower as, in metal-poor environments, less oxygen
is present to form CO, making the CO abundance with respect to H$_2$
lower.
 That would make the total mass-loss rates
higher than what we have determined below.  Thus, the total mass-loss rates
we estimate are  lower limits.

Table~\ref{line_values} shows the properties of the CO lines
observed by \cite{lag10} and the total mass-loss rates they
derived.  One of the star from our sample, IRAS 12560, was 
previously observed in the CO J\,$=$\,2$\rightarrow$\,1 line 
by \cite{gro97}.  They derived a total mass-loss rate of
0.8$\times$10$^{-6}$M$_{\odot}$yr$^{-1}$, which agrees fairly well
with our estimate of 1.3$\times$10$^{-6}$M$_{\odot}$yr$^{-1}$.

Knowing the total mass-loss rate and the DPR allows us to 
estimate the gas-to-dust mass ratio $\psi$.  Assuming that 
the outflow velocities of the dust and gas are the same, 
$\psi$ is simply the ratio of the total mass-loss rate to
the DPR.  Table\,\ref{line_values} presents our estimates for
$\psi$ using the DPRs from the radiative transfer models 
($\psi_D$), and infrared colors ($\psi_c$).

\section{Discussion} 

\subsection{Expansion velocities} 

\begin{figure}
\begin{center}
\includegraphics[width=9cm]{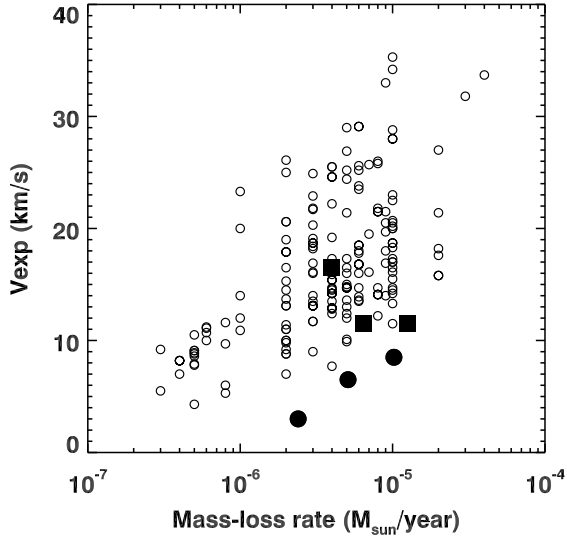}
\caption{\label{mlr_vexp} Expansion velocity (in km\,s$^{-1}$) 
as a function of the total mass-loss rate (in 
M$_{\odot}$yr$^{-1}$).  Open symbols are carbon stars in the
Galactic disc.  Filled symbols are the stars from our sample,
with squares for the thick disc and circles for the Halo.}
\end{center}
\end{figure}

\cite{lag10} showed that the Halo carbon stars they observed 
have a rather high total mass-loss rate but a low expansion 
velocity.  This result agrees with theoretical predictions by
\citep{wac08} that the expansion velocity of carbon stars decreases at lower
metallicity while the mass-loss rate remains stable.  They 
argue that these trends arise from less efficient dust 
formation at low metallicity. As the dust accelerates the 
winds less efficiently at low metallicity, the expansion 
velocities of the envelopes are lower. But only a moderate 
amount of dust is needed to accelerate the wind.  Once this 
threshold is reached, the mass-loss rate will remain 
constant.  Fig. \ref{mlr_vexp} shows the expansion velocity 
as a function of the total mass-loss rates for Galactic disc 
stars (open symbols) and our sample (filled symbols). The 
values for stars in the Galactic disc are from \cite{lou93}.
The observations used for these work were also made in CO.

At a given mass-loss rate, stars from the Halo and the 
Sagittarius Stream have a low expansion velocity.  The star 
with an expansion velocity of 16.5 km\,s$^{-1}$ (IRAS~08427) 
belongs to the thick disc, as well as the two stars with 
expansion velocities of 11.5 km\,s$^{-1}$ (IRAS~04188 and 
IRAS~11308). These stars do not appear as outliers in this 
figure, even though their expansion 
velocities are in the lower part of the observed range.

The stars from the Sagittarius Stream and the Halo are clear 
outliers and have low expansion velocities for their 
mass-loss rates.  This work thus extends the conclusion 
drawn by \cite{lag09}, who studied the DPRs only.
The millimeter CO studies give a picture that somewhat contradicts the conclusions
drawn from the infrared. While the infrared spectral properties are consistent with a metal-rich character,
the low outflow velocities are much more typical of metal-poor stars.

 
\subsection{Gas-to-dust mass ratios} 

Observations of CO with JCMT allow us to measure the total 
mass-loss rates of stars, while infrared observations enable 
us to determine their DPRs.  We can thus estimate their 
gas-to-dust mass ratio $\psi$. 
{As the metallicity of the individual stars we observed is
not known we assume a solar value for the CO abundances, so that
the total mass-loss rates we derived have to be seen as lower limits.
 Thus, the gas-to-dust mass ratios are also lower limits.
The estimated values of $\psi$ range from $\sim$200 (the 
value typically assumed for Galactic stars) to 1300.  
The upper range of our estimates for $\psi$ is high, as 
might be expected for metal-poor stars, where we expect less dust to 
form for a given amount of gas.  IRAS~08427 has a gas-to-dust 
ratio close to the canonical Galactic value of $\sim$200.  This 
star belongs to the thick disc and has the highest expansion 
velocity and the lowest $\psi$ in our sample.  

Some caution is warranted when drawing conclusions from such 
small samples, but these observations are pointing to high
gas-to-dust mass ratios and low expansion velocities in 
carbon stars in the Halo.  To generalize, the gas-to-dust 
mass ratios of carbon stars appear to increase at lower
metallicities.  Such a trend is expected from a theoretical 
point of view \citep{vl00,wac08}, as dust formation is 
expected to be less efficient in metal-poor environments.

IRAS~04188 is an outlier, as it is a thick disc star with a 
high gas-to-dust ratio.  It is interesting to note that the 
spectral energy distribution of this star could not be fitted 
with the same standard parameters as the other stars.  A 
rather cool star (2650~K) was needed for the fit, as well as 
very hot dust (1700~K) for the dust inner radius.

For some of the thick disc stars we observed, $\psi$ is quite  large.
We measured $\psi$ by assuming that the drift velocity (the difference between the dust
and the gas velocity) is equal to zero. 
If the outflow velocities of the gas
  and dust were decoupled, then the actual value for $\psi$
  would be smaller than indicated here. Such a decoupling would require 
less efficient transfer of
  momentum from the radiatively accelerated dust grains to the
  gas in more metal-poor environments, which might be possible
  if the grain-size distribution were shifted to larger grains.  We note that
  the coupling efficiency is determined primarily by the gas density.

\subsection{Metallicity of the Galactic halo carbon stars}
\label{alpha}

\begin{figure}
\begin{center}
\includegraphics[width=9cm]{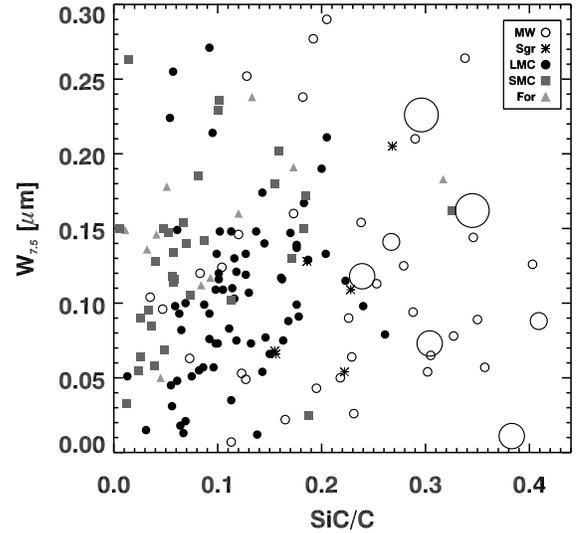}
\caption{\label{sic_7} The equivalent width of the C$_2$H$_2$ 
absorption band at 7.5~\mum\ as a function of the ratio of 
the strength of the SiC dust emission feature at 11.3~\mum\
compared to the underlying continuum.  The symbols are as
defined before, with the largest open circles representing
the Halo stars, and the intermediate open circles 
representing stars in the thick disc.}
\end{center}
\end{figure}

Fig.\,\ref{sic_7} displays the equivalent width of the 
7.5~\mum\ absorption band from C$_2$H$_2$ as a function of 
the strength of the SiC dust emission feature.  \cite{lag09}
introduced this plot and showed that it can diagnose the
metallicity of carbon stars observed spectroscopically in the 
infrared.  Despite some scatter and mixing between the 
populations, the carbon stars from the Milky Way are clearly 
concentrated toward the lower right corner.  Moving 
progressively up and to the left, one moves through the 
region dominated by the LMC, followed by the SMC.  The Fornax 
stars trace the same region as the SMC, consistent with the 
similar metallicities of the carbon stars sampled with the 
IRS to those in the Magellanic Clouds, as described by 
\cite{slo12}. {The position of stars on this diagram depends also on its initial
mass, sp that it can be use only as an indication of the metallicity in the present case.

The carbon stars from our sample, both in the thick disc and
Halo, as well as our two comparison stars in the foreground
of the Sgr dSph and the one star near Lyng{\aa}~7~V1, occupy 
the same region as the Galactic carbon stars.

As mentioned in Section\,\ref{spectra}, decreasing the initial metallicity
tends to lead to a higher C/O ratio and thus stronger C$_2$H$_2$. This would
also decrease the initial abundance of Si and thus decrease the SiC strength.
But silicon, being a heavy $\alpha$-element, it is not synthetized in AGB
stars. The location of the star on the diagram Fig.\,\ref{sic_7} thus not only
depends on [Fe/H] but also on the initial $\alpha$-element abundance.

{On the other hand, observations of C-rich post-AGB stars with the 21 microns feature
 in the Galaxy and the Magellanic Clouds (Decin et al., 1998; De Smedt et al., 2012 and references therein)
indicate that metal-poor carbon stars can have significantly enhanced C, O and Si abundances.
Such an enhancement would explain the strong SiC feature observed in the metal poor carbon stars we observed in the halo.
 However, it is not clear why metal-poor carbon stars would produce more Si in the halo than in other galaxies of the Local Group, as we have shown
AGB stars in the SMC, LMC, Fornax and Sculptor had a weaker SiC feature than Galactic stars.

Enrichment of the ISM by Type II supernovae, which are known
  producers of $\alpha$ elements is a tempting explanation, but the
  known behavior of [$\alpha$/Fe] as a function of [Fe/H], as
  illustrated by \citet[][see their Fig.\ 11]{tol09} is not 
  consistent with our spectra. The [$\alpha$/Fe] abundance 
  increases by $\sim$one decade while [Fe/H] drops by two
  decades.  In terms of absolute abundance, [$\alpha$/H] is
  decreasing at lower metallicities. The bottom panels of Fig.~2
  contradict this expected behavior. They show that the SiC 
  features in our Halo and thick disc spectra are strong
  compared to most of the other spectra in {\it absolute terms}
  and not just as a ratio to amorphous carbon, as shown in the
  upper panels.

The two Halo stars observed with Spitzer and JCMT appear to have a low expansion velocity (an indication that they are metal-poor)
and a very strong SiC feature. 
Those stars could thus either be similar to Galactic disc stars or metal-poor and  $\alpha$-element enhanced.

\subsection{Origin of the Galactic halo carbon stars}
The typical age of halo stars is 10 Gyr, so no AGB stars more massive than 1
M$_{\odot}$ formed in the Halo should be observable now. Models of AGB stars with such a
low mass do not dredge up enough carbon to make the stars  carbon-rich.
 But carbon-rich stars are observed in the halo (Totten \& Irwin,
1998; Mauron, 2008).   The fact that we observe carbon stars in the halo could be a clue that
these models are not right, or that these stars were formed outside of the halo.
{ Comparing the luminosity of the star we observed with Fig. 20 from (Vassiliadis \& Woods 1993)
indicates that the initial masses of the stars in our sample are between 2 and 3M$_{\odot}$.
This is one more indication that those stars might have formed outside the Galactic halo.
We propose two possible external origins for these halo carbon
stars.

It has been proposed that carbon AGB stars in the halo could originate from
the tidal disruption of Sgr dSph Galaxy, orbiting the Milky Way with a period
of about 1 Gyr (Ibata et al.\, 2001). Their model of the Sgr dSph stream shows
a good agreement between the orbit of the stream and those of observed carbon
stars.

Out of the three halo stars we observed with the JCMT, one, IRAS 12560,
likely belongs to the Sagittarius stream (Lagadec et al.\, 2010) and thus
 originates from Sgr dSph.  These three Halo stars have a very low
envelope expansion velocity, indicative of a  metal-poor nature
(Lagadec et al.\, 2010).  The Spitzer spectra we present here show that those
stars present strong C$_2$H$_2$ absorption bands, as observed in metal-poor
carbon stars (see e.g. Sloan et al.\, 2009). These two arguments strongly
support the hypothesis that these stars are metal-poor.  As shown in
Section \ref{alpha}, the halo carbon stars we observed could be enhanced in
$\alpha$-elements. They would thus have formed in a burst-like stellar
population. As they are carbon stars, this population is necessarly younger
than the halo itself.  The Sgr dSph have such a stellar population. The halo
carbon stars we observed may thus come from the Sgr dSph galaxy or a similar
galaxy orbiting the Milky Way. The fact that the dust properties of the Sgr dSph
carbon stars are very similar (Lagadec et al.\, 2009) supports this
argument. This galaxy contains a metal-rich population (Kniazev et al. 2007)
but it is not known whether this population is represented in the tidal tails.

But those stars could also be similar to Galactic disc stars (Section\,\ref{alpha}).
This leads naturally to the question of how stars
formed near the Galactic plane could wind up several kpc 
above or below it. They could be runaways from
the thin disc.  One interesting scenario is that they may 
have been ejected from the break-up of a binary, possibly as 
a result of the interaction of a binary and a single star, 
as suggested for more massive runaways \citep[e.g.,][]{hoo01,
fuj11}. While carbon stars are bright and easily
detected, they represent perhaps one thousandth of a star's
total lifetime.   This would
point to a larger population of less evolved 
intermediate-mass stars now in the Halo, but relatively
metal-rich.

To discriminate the two possibilities, we would need to obtain measurements of [Fe/H]
and different $\alpha$-elements abundances for the stars we observed. {Such measurements are made difficult
by the fact that those stars are very faint in the optical.

A final possibilty is that these are extrinsic carbon stars,  with the carbon
enrichnent caused by mass transfer from a more massive, faster evolving
companion. 


 \section{Conclusions}

We presented {\it Spitzer} IRS spectra of four Galactic disc and halo carbon stars.
 Typical features of carbon stars with SiC dust 
emission and C$_2$H$_2$ molecular absorption are seen  in the four spectra.
  Dust radiative 
transfer models enabled us to determine the dust-production 
rates for these stars.  We measured the gas mass-loss rates 
and expansion velocities from the stars via modeling of the 
CO J $= 3 \rightarrow 2$ transition.  The gas-to-dust mass 
ratio $\psi$ was then measured by simply dividing the gas and 
dust mass-loss rates. The measured mass-loss rates and $\psi$
are in the range 1--4\,10$^{-6}$M$_{\odot}$yr$^{-1}$ and 
200--1300 respectively.  The stars in the halo have a low 
expansion velocity (around 7 km/s), strong C$_2$H$_2$ 
molecular absorption bands,  strong SiC emission and a 
 high gas-to-dust mass ratios. 

The strong SiC emission we observe could be due to an
overabundance of $\alpha$-elements (like silicon) in the 
Halo, that these stars produce silicon or to the fact that those stars metallicities similar
to the Galactic disc.  In the latter case, they may have
escaped from the disc.  The low expansion velocities
and strong C$_2$H$_2$ absorption bands are arguments in
favor of the former.  However, explaining the $\alpha$ 
enhancement in the Halo is a challenge.  We conclude that
these spectra present us with some intriguing contradictions, 
with the dust properties being similar to metal-rich carbon stars and 
the gas properties similar to metal-poor carbon stars.

{Those AGB stars could have formed outside of the Halo,
as the mass of these carbon stars is inconsistent with a formation in the Halo.
The halo carbon  AGB stars we 
observed could thus have two possible origins.
They could have  formed in a galaxy similar 
to Sgr dSph, orbiting the Milky Way and being gradually 
stripped by tidal forces or they could have escaped from the Galactic disc.

 This work has shown the necessity 
to combine mid-infrared spectroscopy with millimeter CO 
observations to fully characterize the dust production {\it and} 
gas mass-loss from AGB stars.  Observations of metal-poor AGB 
stars in Local Group dwarf galaxies are necessary to study 
the dependence of this mass-loss on metallicity. The Atacama 
Large Millimeter/submillimeter Array (ALMA) will allow such 
observations.  Finally near-infrared spectroscopy of the 
individual stars will be necessary to precisely determine the 
metallicity and $\alpha$-element abundances  of the observed stars.



\section*{Acknowledgments}
The research leading to these results has received funding from the European Community's Seventh Framework Programme (/FP7/2007-2013/) under grant agreement No 229517. 
These observations were obtained as guaranteed time for the 
IRS team at Cornell, who are supported by by NASA through 
Contract Number 1257184 issued by the Jet Propulsion Laboratory,
California Institute of Technology under NASA contract 1407.
The research in this paper has benefited from NASA's 
Astrophysics Data System, and the SIMBAD and VIZIER databases, 
operated at the Centre de Donn\'{e}es astronomiques de 
Strasbourg.

\label{lastpage}

\end{document}